\documentclass[aps,prb,twocolumn,amsmath,amssymb,nofootinbib,
  superscriptaddress]{revtex4-2}

\usepackage{graphicx}
\usepackage{bm}
\usepackage[colorlinks]{hyperref}
\usepackage[usenames,svgnames]{xcolor}
\usepackage{makecell}
\usepackage[normalem]{ulem}

\newcommand{\Hop}{H}

\newcommand{\distance}{d}
\newcommand{\sgx}{\sigma^x}
\newcommand{\sgy}{\sigma^y}
\newcommand{\sgz}{\sigma^z}
\newcommand{\nproc}{n_{{\rm proc}}}

\newcommand{\norm}[1]{{\| #1 \|}}  
  
\newcommand{\ket}[1]{{ |{#1} \rangle }}  
\newcommand{\bra}[1]{{ \langle {#1} | }}
\newcommand{\av}[1]{{ \langle {#1} \rangle }}
\newcommand{\braket}[2]{{ \langle {#1} | {#2} \rangle}}

\newcommand{\cc}[1]{\mbox{~\cite{#1}}}
\newcommand{\cRef}[1]{Ref.\cc{#1}}
\newcommand{\cRefs}[1]{Refs.\cc{#1}}
\newcommand{\Fig}[1]{Fig.~\ref{#1}}
\newcommand{\Eq}[1]{Eq.~\eqref{#1}}

\newcommand{\EqDef}{\stackrel{\mathrm{def}}{=}}

\DeclareMathOperator*{\Tr}{Tr}

\newcommand{\smt}{Science, Mathematics and Technology Cluster, Singapore University of Technology and Design, 8 Somapah Road, 487372 Singapore} 
\newcommand{\epd}{Engineering Product Development Pillar, Singapore University of Technology and Design, 8 Somapah Road, 487372 Singapore} 
\newcommand{\cqt}{Centre for Quantum Technologies, National University of Singapore 117543, Singapore} 

\newcommand{\hnu}{Key Laboratory of Low-Dimensional Quantum Structures and Quantum Control of Ministry of Education, Department of Physics and Synergetic Innovation Center for Quantum Effects and Applications, Hunan Normal University, Changsha 410081, China
}

\begin{document}

\title{Block belief propagation algorithm for two-dimensional tensor networks}

\author{Chu Guo}
\affiliation{\hnu}

\author{Dario Poletti} 
\affiliation{\smt} 
\affiliation{\epd} 
\affiliation{\cqt} 
\affiliation{MajuLab, CNRS-UNS-NUS-NTU International Joint Research Unit, UMI 3654, Singapore}

\author{Itai Arad} 
\email{arad.itai@fastmail.com}
\affiliation{Faculty of Physics, Technion, Haifa 3200003, Israel}



\begin{abstract}
  Belief propagation is a well-studied algorithm for approximating
  local marginals of multivariate probability distribution over
  complex networks, while tensor network states are powerful tools
  for quantum and classical many-body problems. Building on a 
  recent connection between the belief propagation algorithm and the
  problem of tensor network contraction, we propose a block belief
  propagation algorithm for contracting two-dimensional tensor
  networks and approximating the ground state of $2D$ systems.  The
  advantages of our method are three-fold: 1) the same algorithm
  works for both finite and infinite systems; 2) it allows natural
  and efficient parallelization; 3) given its flexibility it would
  allow to deal with different unit cells. As applications, we use
  our algorithm to study the $2D$ Heisenberg and transverse Ising
  models, and show that the accuracy of the method is on par with
  state-of-the-art results.
\end{abstract}

\maketitle

%
%
\section{Introduction}

Two-dimensional quantum systems represent an important class of
problems of long-lasting theoretical and practical interest.  Many
such systems are notoriously difficult to investigate numerically,
especially in the strongly correlated regime. For this reason,
developing efficient and accurate numerical methods is always in
need.  Quantum Monte Carlo\cc{FoulkesRajagopal2001} and tensor
network methods\cc{Orus2014} have both been demonstrated to be
successful approaches for 2D systems in the past decades. 
However,
the quantum Monte Carlo methods often
suffer from the sign problem\cc{TroyerWiese2005}.
Projected
entangled pair states (PEPSs) are a class of two-dimensional tensor
networks (TNs)\cc{VerstraeteCirac2004b, Orus2014, Orus2019,
CiracVerstraete2021} that can be used to study both finite (fPEPS) and infinite (iPEPS) 2D systems. In particular, the iPEPS
algorithms have attracted great attention during the last twenty
years because of their lack of a sign problem and their ability to
deal with strongly-correlated systems\cc{CorbozTroyer2014,
CorbozMila2013, CorbozMila2014, ZhengChan2017, LiaoXiang2017,
LiXie2022, XieXiang2012}.  A key challenge for PEPS algorithms is an
efficient and stable procedure for computing 
expectation values of local observables, which can become a
$\#$P-hard problem in worst cases in 2D and higher
dimensions\cc{SchuchCirac2007}. In a number of physical scenarios
the problem tends to be more tractable, but it still requires high
computational resources and advanced numerical schemes.  Existing
approaches include, for example, the boundary MPS (bMPS)
method\cc{VerstraeteCirac2004b, VerstraeteMurg2008}, the
corner-transfer matrix (CTM) method \cc{NishinoOkunishi1996,
OrusVidal2009}, Monte-Carlo sampling\cc{ref:SV2007-MC-TN,
ref:WPV2011-MC-TN}, and variants of tensor network renormalization
schemes\cc{LevinNave2007, JiangXiang2008, EvenblyVidal2015}.

Borrowing ideas
from classical probabilistic models, we propose an alternative method for (approximately) contracting two-dimensional TNs, which is highly parallelizable and is
flexible to deal with both finite and infinite 2D systems.
There is a close similarity between quantum and classical many-body
systems and multivariate probabilistic models. In both cases, the
states live in spaces of exponential dimension, and it is important
to (approximately) compute local properties of the system: local
expectation values in the former cases, and local marginals in the
latter case\cc{WainwrightJordan2008}. Therefore, methods developed
in one field could often benefit the other. Belief propagation (BP)
is a well-established statistical inference algorithm for computing
local marginals of multivariate probabilistic models on complex
networks, which is known to be exact for tree
networks\cc{Pearl1982}. BP has been widely applied to very diverse
areas, ranging from Bayesian inference\cc{Pearl1982}, statistics
physics\cc{MezardVirasoro1987, ToonMendes2011, KarrerZdeborova2014},
combinatorial optimizations\cc{MezardZecchina2002,
MezardMontanari2009}, and epidemic spreading\cc{KarrerNewman2010},
to name a few. For networks that contain a large number of loops or
long range correlations, BP could result in poor performance since
it only considers the direct neighbors of each node and treats the
environment of a node in a separable way. This shortcoming has been
the subject of an extensive research that resulted in a plethora of
algorithms that generalize and improve BP in
different ways\cc{YedidiaFreemanWeiss2000, YedidiaFreemanWeiss2005,
ref:Pelizzola2005-CVM, ref:Montanari2005-LCB, ref:Chertkov2006-LCB,
ref:Mooij2007-LCB, ref:LMRR2013-replica, WangZhou2013,
ZhouZheng2015, ref:CantwellNewman2019-LoopyBP,
ref:KirkleyNewman2021-LoopyBP}. Broadly speaking, these algorithms
clump together neighboring nodes and treat small loops exactly, at
the expense of higher computational resources and elaborated
bookkeeping. 

A first attempt to use BP to approximate TN contraction was
presented in \cRefs{AlkabetzArad2021, SahuSwingle2022}. When applied
with the imaginary time evolution (ITE) to PEPS, it was shown to be
equivalent to the simple-update algorithm\cc{AlkabetzArad2021}, in
which the environment is approximated in a separable, mean-field
way\cc{JiangXiang2008}.  In this work we present a generalization of
the BP algorithm of \cRef{AlkabetzArad2021}, which we call
\emph{block belief propagation} (blockBP). The algorithm partitions
the system into non-overlapping blocks of spins, and uses the
framework of \cRef{AlkabetzArad2021} to define BP messages between
these blocks. From the converged messages one can obtain an
approximation of the local environment of every block, which is far
superior to the mean-field environment of the original BP, since the
loops within each block are accounted for.  We benchmark our
algorithm on the 2D Heisenberg model and transverse Ising model,
where we show that it can reach similar precision to current
state-of-the-art methods. Importantly, we show that our
approach can be easily parallelized, thus leading to a much higher
computational efficiency compared to other tensor-network-based
state-of-the-art methods.  Furthermore, we also demonstrate that the
same algorithm can be readily used both for finite systems and
infinite systems with translational invariance. Lastly, the
algorithm can be applied to different unit cells and systems with
different geometries. As such, blockBP might also be potentially
used in quantum chemistry problems with less regular
structures\cc{molecular_electronic_structure_theory}.

\begin{figure}
  \includegraphics[scale=1]{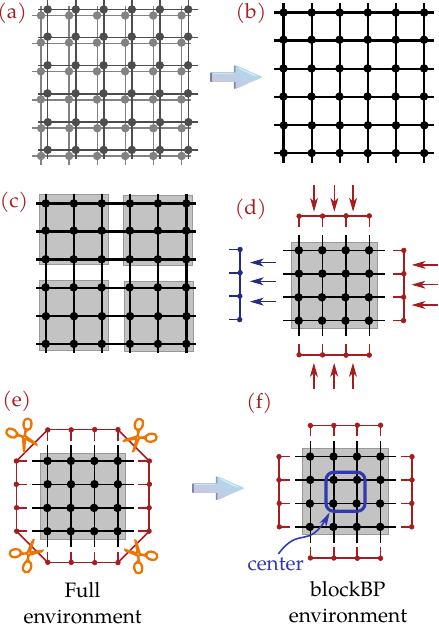}
    \caption{(a,b) Forming a double layer 2D PEPS by
    contracting the physical legs of $\ket{\psi}$ and $\bra{\psi}$.
    (c) Partitioning of the system into blocks (grey regions).  (d)
    Computing the left output MPS message for a given block. (e) The
    exact block environment. (f) The blockBP environment, which
    approximates the exact block environment using a product of the
    converged incoming MPS messages. These can be used to compute
    the local environments inside the center of the block.}
    \label{fig:blockBP-sketch}
\end{figure}

%
%

\section{The blockBP algorithm}

To present the blockBP algorithm, we
assume basic familiarity with TN such as MPS and
PEPS\cc{Schollwock2011, Orus2014}. More information for the
unfamiliar reader can be found in Appendix.~\ref{sec:zipup} and Appendix.~\ref{sec:ite}. Consider a 2D many-body whose state is given by
a PEPS $\ket{\psi} = \sum_{i_1, \ldots, i_n}\Tr(T^{i_1}_1\cdots
T_n^{i_n})\ket{i_1, \ldots, i_n}$. Here, $T_1, \ldots, T_n$
represent the local tensors of the PEPS, with dimension of the
virtual legs, also known as bond dimension, $D$, and $i_1, \ldots,
i_n$ denote the physical legs. Virtual legs are omitted for brevity,
and $\Tr(\cdots)$ denotes TN contraction.


The blockBP algorithm is an efficient
method to approximate the local environments in a 2D PEPS.  The
first step of the algorithm is to move to a double-layer PEPS. This
is a closed TN that represents the scalar $\braket{\psi}{\psi}$ by
contracting the physical legs between a bra TN $\bra{\psi}$ and a
ket TN $\ket{\psi}$ --- see \Fig{fig:blockBP-sketch}(a,b). Next, we
partition the system into non-overlapping blocks as in
\Fig{fig:blockBP-sketch}(c). The goal of the blockBP algorithm is to
simultaneously find an approximation to the environment of each
block. The exact environment can be written as a periodic MPS that
surrounds the block (\Fig{fig:blockBP-sketch}(e)), which is equal to
the contraction of the entire doubled-layered PEPS \emph{outside}
the block. Knowing the environment of each block allows an easy
calculation of the expectation values within each block, as well as
an optimal update of the tensors of the block during an ITE or
variational algorithms. However, calculating it directly might be a
daunting task when the system size is large.

The key approximation of the blockBP algorithm is to break the block
environment into a product of MPSs, where each MPS belongs to a
block edge corresponding to an adjacent block (see
\Fig{fig:blockBP-sketch}(e,f)). This inflicts an unavoidable error
on the environment, as it breaks some of its entanglement, but in
many cases this error is negligible (see discussion on this point
below).

The MPS states that make each environment are calculated in a
self-consistent way using a BP algorithm. We define a
message-passing algorithm in which at step $\ell$ each block
receives (sends) MPS messages from (to) its adjacent blocks along
their common edge. We denote by $m^{(\ell)}_{A\to B}$ the MPS
message from block $A$ to block $B$ at the $\ell$'th iteration. Let
$T_A$ denote the part of the double layer PEPS that corresponds to
block $A$. This is an open TN with double-layer legs at each edge of
the block. The MPS message from block $A$ to block $B$ at the
$\ell+1$ iteration is obtained by contracting $T_A$ with all
incoming MPS messages to block $A$ of the $\ell^\text{'th}$
iteration, except for the message from $B$ (see
\Fig{fig:blockBP-sketch}(d)):
\begin{align} 
\label{eq:blockbp-messages}
  m_{A\rightarrow B}^{(\ell+1)} =
    \Tr \left(T_A \prod_{A'\in N_A \backslash \{B\}} 
      m^{(\ell)}_{A'\rightarrow A} \right).
\end{align}
When the underlying blocks graph is a tree, then just as in ordinary
BP, the fixed point of \Eq{eq:blockbp-messages} describes the exact
environment of each block. Note that indeed in such case, the
environment of each block breaks into a product of MPSs. 

The iteration in \Eq{eq:blockbp-messages} can be done efficiently
using, e.g., the bMPS method. In addition, messages of different
blocks can be computed in parallel. We stop the iterations when the
average distance between the normalized MPS messages at two
consecutive iterations is sufficiently small (say, smaller then
$\epsilon=10^{-5}$). This usually happens in less than 10 steps.

For blocks of size $1$, our algorithm reduces to the BP contraction
algorithm of \cRef{AlkabetzArad2021}, which has been shown to be
equivalent to the simple-update algorithm. These algorithms are
exact on tree-like (i.e., loop free) TNs, but may fail on more
complicated graphs. In contrast, for larger block sizes blockBP can
yield very accurate correlations inside a block, as it considers
exactly the loops within it.
We note, however, that as the
entanglement around every block is broken between every two adjacent
block edges, some inaccuracies are unavoidable.
In non-critical
systems, this problem can be largely mitigated if we consider only
observables in the \emph{center} of a block (see
\Fig{fig:blockBP-sketch}e). For these sites, most of the
correlations in the local environment come from near-by tensors,
which lie inside the block and are therefore properly accounted for.
By increasing the block size, the accuracy for the center sites can
be improved at the expense of higher computational costs. In this
respect, our algorithm shares some similarity with the
cluster-update algorithm\cc{LubaschBanuls2014}, where the local
environment of a row $i$ is calculated by contracting a mean-field
environment of the rows at distance $\delta$ from $i$, together with
the full TN of the rows $i-\delta, \ldots, i+\delta$ that surrounds
$i$, although we do not need to rely on a mean-field environment. 

The relation between the block size, the system size and the center
size can be chosen differently depending on the specific tasks to
balance the computational efficiency and the accuracy.  In addition,
since we only calculate the local environment of spins at the center
of each block, we need several partitions of the system into blocks
in order to cover all the bonds (neighboring spins) in the system. 
Bookkeeping the different partitions becomes much easier if we work
with periodic boundary conditions (PBC), in which different
partitions correspond to shifting the lattice in different
directions. This technique is useful also in the case of open
boundary conditions (OBC), as we can embed an OBC PEPS in a PBC PEPS
by setting the bond dimension at the boundaries to $1$. Furthermore,
we can easily adjust it for an infinite, translational invariant
PEPS, described by tensors in a unit cell (iPEPS). In such case, one
block is sufficient to calculate the environment of a unit cell at
its center. The block sends and receives messages to itself, which
mimics the case where it is surrounded by its own copies.

To efficiently perform the contractions in
\eqref{eq:blockbp-messages}, we use the bMPS method with a bond
dimension $\chi_m$, which we typically set to $\chi_m=D^2$. Once the
messages have converged, we use them to calculate the local
environment of spins in the center of a block, so as to compute
observables or optimize the tensors in the center to approach the
ground state. To this aim, we use the bMPS algorithm once more, this
time with a larger bond dimension $\chi$, which we set to
$\chi=2D^2+10$ if not specified otherwise.

In this paper, in order to obtain the ground state, we use blockBP
in combination with ITE, and we refer to this as blockBP update.
More details can be found in the Appendix.~\ref{sec:ite}. Note that
in our algorithm we make use of bMPS, but only over small portions
of the system, our blocks, and not on the entirety of the system as
in the bMPS algorithm.

Before presenting our numerical results, we summarize the important
features of blockBP: 1) it works for both finite and infinite 2D
systems, as \Eq{eq:blockbp-messages} is agnostic about the
boundaries, and the differences between them are automatically
revealed when computing messages.  For finite systems, whenever the
boundary is met, the information before the boundary will be lost
because the virtual dimension of the tensor at the boundary is $1$.
Without a boundary, instead, the fixed point will directly
correspond to the infinite limit.  2) the overall complexity of
blockBP is that of bMPS times a constant factor due to the overhead
of computing the messages. However, \Eq{eq:blockbp-messages} can be
evaluated independently for each block, and after the messages have
converged, the calculation of local environments inside the center
is also completely independent. Since these are the most
computation-intensive parts of the algorithm, blockBP can be
efficiently parallelized on distributed architectures; 3) blockBP is
flexible on the choice of the unit cells (size and geometry), which
is particularly useful for infinite systems with large unit cells.
In the following, we demonstrate the efficiency and accuracy of
blockBP for both finite and infinite 2D systems. 


%
%

\section{blockBP for finite systems}

We first demonstrate the
accuracy of blockBP by applying it to compute the ground state of
the finite-size transverse Ising (TI) model, whose Hamiltonian is
$\Hop_{TI} \EqDef -\sum_{\av{i,j}}\sgz_i\sgz_j - B\sum_k\sgx_k$ with
$B$ the strength of the transverse field, and anti-ferromagnetic
Heisenberg (AFH) model with Hamiltonian $\Hop_{AFH} \EqDef
\sum_{\av{i, j}} (\sgx_i\sgx_j + \sgy_i\sgy_j + \sgz_i\sgz_j)/4$.
Both models are defined on a square lattice with nearest-neighbors
interactions. For the ITE of finite systems, we chose the center
size to be the same as the block size for efficient update. The
final energies for the finite systems are computed using bMPS. The
energies per site of the final PEPS are shown in
Table.~\ref{tab:ising_obc} for TI and in Table.~\ref{tab:xxx_obc}
for AFH, which are compared to the bMPS full-update results from
Ref.\cc{LubaschBanuls2014b}.  We see that the blockBP energies are
very accurate for all cases.  For $D=6$ the blockBP energies are
slightly larger, which is likely because a larger block size should
be used to account for longer range correlations. We use at most
$D=6$ due to the unfavorable scaling of the algorithm with $D$,
namely $O(D^{12})$, essentially because we use bMPS for each block.

\begin{table}[!htb]
  \begin{center}
    \caption{Ground state energies of the $21\times 21$ transverse 
    Ising model with $B=2.5, 3, 3.5$. We used a block size of
    $7\times 7$ ($9$ blocks in total) for blockBP. }
    \label{tab:ising_obc}
    \begin{tabular}{c|c|ccc}
    \hline
    \hline 
    & $D$ &  $2.5$ & $3$ & $3.5$  \\
    \hline 
    & $2$   & $-2.77340(2)$ & $-3.18128(6)$ & $-3.64849(3)$  \\
    bMPS\cc{LubaschBanuls2014b} & $3$   & $-2.77346(1)$ & $-3.18242(1)$ & $-3.64873(1)$  \\
    & $4$   & $-2.77346(1)$ & $-3.18243(1)$ & $-3.64873(1)$  \\
    \hline 
    & $2$   & $-2.77334(7)$ & $-3.18121(9)$ & $-3.64842(7)$  \\
    blockBP & $3$   & $-2.77346(3)$ & $-3.18242(3)$ & $-3.64872(9)$  \\
    & $4$   & $-2.77346(4)$ & $-3.18244(4)$ & $-3.64873(1)$  \\
\hline
    \end{tabular}
  \end{center}
\end{table}

\begin{table}[!htb]
  \begin{center}
    \caption{Ground state energies of the AFH model of sizes
      $10\times 10$ and $14\times 14$, for which We used block sizes
      of $5\times 5$ and $7\times 7$ respectively.}
    \label{tab:xxx_obc}
    \begin{tabular}{c|cc|cccc}
    \hline
    \hline
     & $10\times 10$ & & $14\times 14$ & \\
    \hline 
    $D$ & bMPS\cc{LubaschBanuls2014b} & blockBP & bMPS\cc{LubaschBanuls2014b} & blockBP \\
    \hline
    $2$ & $-0.61310(2)$  & $-0.61310(1)$ & $-0.62631(1)$ & $-0.62631(3)$ \\
    $3$ & $-0.61999(1)$ & $-0.62012(0)$ & $-0.63246(1)$ & $-0.63273(4)$ \\
    $4$ & $-0.62637(2)$  & $-0.62636(2)$ & $-0.63832(3)$ & $-0.63831(8)$\\
    $5$ & $-0.62739(1)$  & $-0.62737(7)$ & $-0.63901(1)$ & $-0.63905(2)$ \\
    $6$ & $-0.62774(1)$  & $-0.62759(3)$ & $-0.63930(1)$ & $-0.63928(3)$ \\
  \hline
    \end{tabular}
  \end{center}
\end{table}

The accuracy of PEPS algorithms rely on the accuracy of the computed
local environment of each bond. To study this, we used blockBP to
compute the reduced density matrices (RDMs) in the ground state of
the TI model at $B=2.5,3.0,3.5$, and compared it to the RDMs that
were calculated using bMPS on the \emph{same} PEPS. We used a system
of $21\times 21$ spins with OBC, where the ground state was
approximated by PEPS with $D=2,3,4$, and was calculated using the
bMPS full-update algorithm. For blockBP, we used a block size of
$7\times 7$ with a center of $2\times 2$. To compare the blockBP and
bMPS RDMs we calculated the trace distance $D(\rho,\sigma)\EqDef
\frac{1}{2}\norm{\rho-\sigma}_1$ of $2$-local RDMs of the horizontal
bonds.  For $B=2.5,3.5$, which are away from the critical value
($B_c\approx 3.044$\cc{BloteDeng2002}), the RDMs computed by these
two methods were very close to each other with average distance
lower than $10^{-5}$ for all the $D$s considered, while for $B=3.0$
we got, on average, a trace distance of $\approx 10^{-3}$ (see
Appendix.~\ref{sec:rdm} for details on this, and Appendix.~\ref{sec:convergence} for further
discussion on convergence).  Interestingly, although the accuracy of
the RDMs computed using our blockBP near the critical point is lower
by about two orders of magnitude than the accuracy away from
criticality, the ground state energy of the resultant PEPS using
blockBP update is as good as those from the bMPS full-update
(Table.~\ref{tab:ising_obc}). 

%
%

\section{Parallel computation}

\begin{figure}
  \includegraphics[width=0.8\columnwidth]{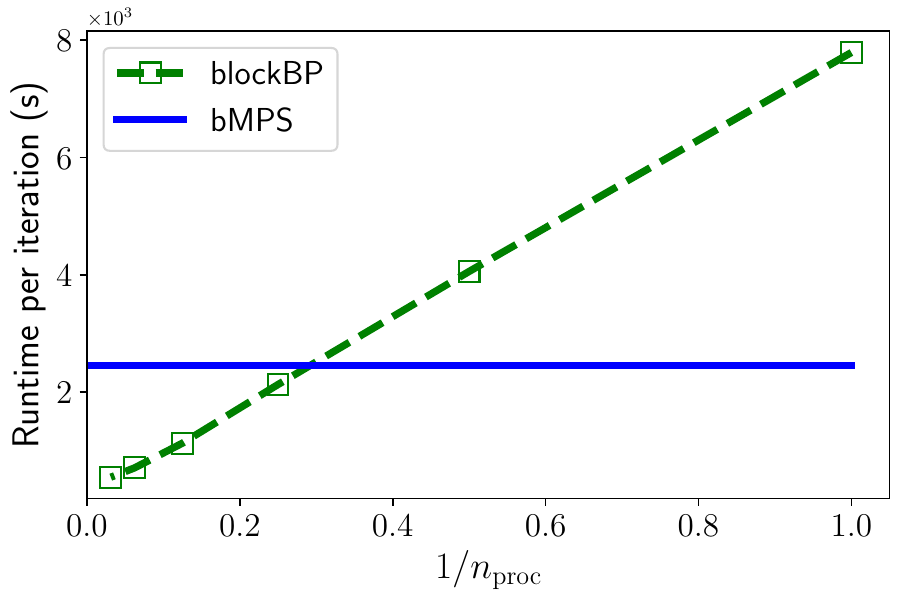}
  \caption{Scaling of the blockBP runtime for an imaginary time
  evolution step versus the number of processes $n_{\rm proc}$, and
  for the AFH model of size $40\times 40$, using a block size
  $5\times 5$. The solid line shows the corresponding runtime of
  bMPS using a single thread for comparison.  } 
\label{fig:fig3}
\end{figure}

In \Fig{fig:fig3} we show the scaling
of the runtime per ITE step (iteration) for blockBP with block size
$5\times 5$ against the number of processes ($\nproc$), for the
$40\times 40$ AFH model. 
We can clearly see that the runtime of
blockBP scales almost linearly as $1/\nproc$ as $\nproc$ increases
up to $\nproc=32$. 
Instead, bMPS acts sequentially on the whole system, and for this reason no approach to parallelize it has, to the best of our knowledge, been found.
For $\nproc=1,2$,
blockBP is slower than bMPS due to the overhead of computing the
messages, but beyond these values of $\nproc$ we see a clear
advantage of using blockBP.

\section{blockBP for infinite systems}

\begin{figure}
  \includegraphics[width=0.9\columnwidth]{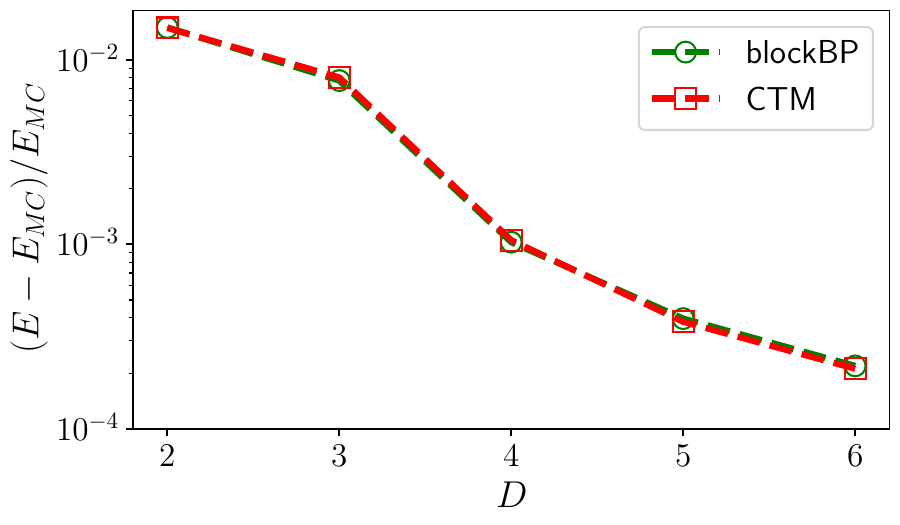}
  \caption{Relative energy compared to the exact energy $E_{MC}$
    obtained by quantum Monte Carlo\cc{Sandvik1997} for the
    infinite AFH model. The green dashed line with circle shows the
    results computed using BlockBP while the red dashed line with
    square shows the CTM results\cc{PhienOrus2015}.  }
\label{fig:fig4}
\end{figure}

Finally, we apply blockBP to study infinite quantum and classical systems.
We first compute the ground state of the infinite AFH model as an iPEPS, for
which we chose a unit cell size as $2\times 2$ and a block size of
$4\times 4$. The final energies are computed by embedding the
$2\times 2$ iPEPS into a $52\times 52$ finite PEPS with randomly
initialized boundaries, and then computing the energy in the center
using bMPS. We have verified the correctness of our energy estimate
by checking that the resultant energies have well converged against
the finite PEPS size. From the results in \Fig{fig:fig4}, we can see
that our precision is on par with the CTM results from
\cRef{PhienOrus2015}, while only requiring a computational
complexity that is similar to the bMPS algorithm applied to just a
$4\times 4$ system. 

\begin{figure}
  \includegraphics[width=\columnwidth]{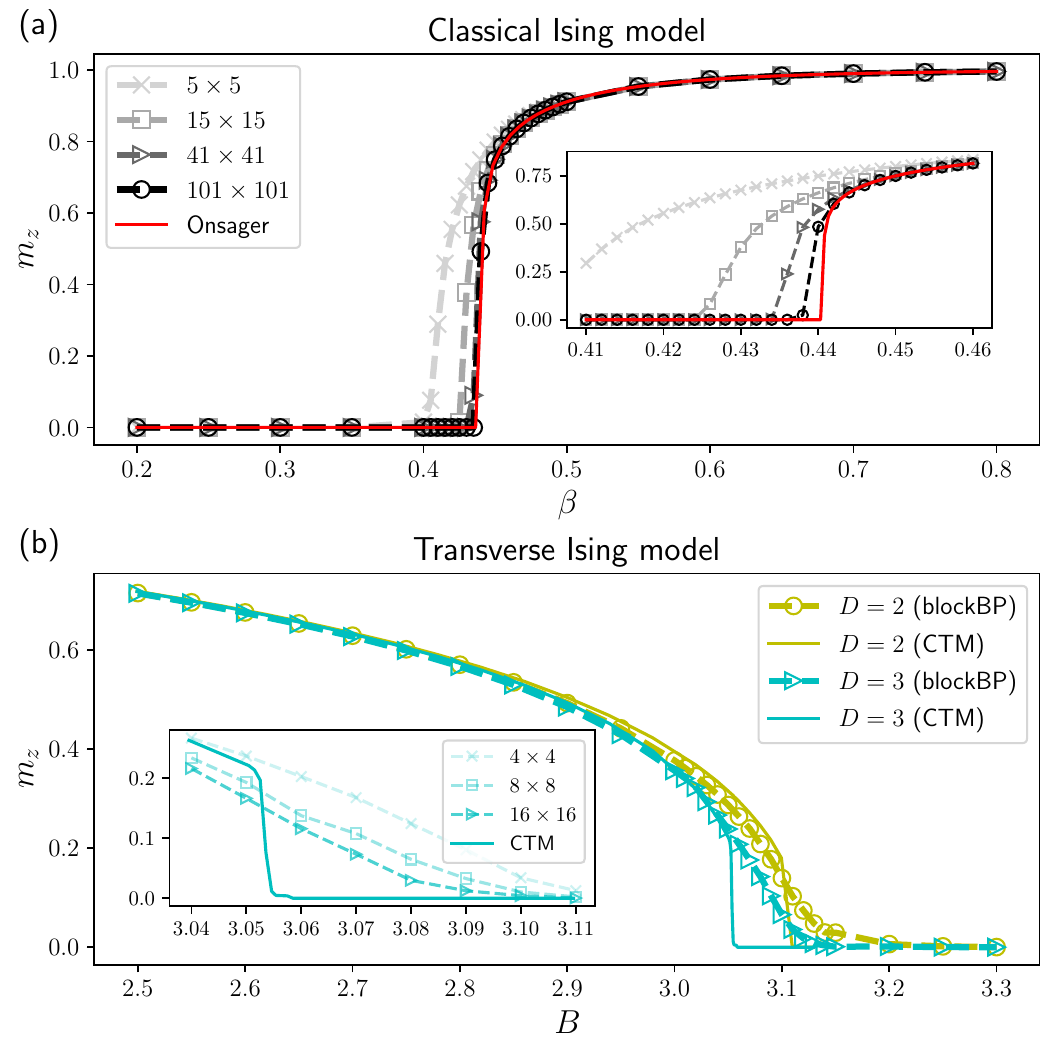}
    \caption{(a) Local magnetization $m_z$ as a function of the
    inverse temperature $\beta$ for the infinite classical Ising
    model. The red solid line is Onsager's exact solution. (b) $m_z$
    as a function of $B$ for the infinite quantum Ising model for
    $D=2$ (orange dashed line with circle) and $D=3$ (cyan dashed
    line with triangle), where we have used a $2\times 2$ unit cell
    and a block size $4\times 4$. The corresponding solid lines are
    results from \cRef{PhienOrus2015}. The inset shows $m_z$
    computed using larger block sizes near the critical point for
    $D=3$.} \label{fig:figS4}
\end{figure}

In \Fig{fig:figS4} we further study the infinite classical Ising model and
the transverse Ising model to demonstrate the versatility of
blockBP.  In \Fig{fig:figS4}(a) we look at the local magnetization $m_z$ of
the Gibbs state of the 2D classical Ising model on a square lattice
(with Hamiltonian $H = \sum_{\av{i,j}}\sgz_i\sgz_j$) as a function
of the inverse temperature $\beta$. This can be written exactly as a
tensor network with $\chi=2$, which is an ideal test ground for
computing local observables since it does not require updating. In
this case we used infinite blockBP to calculate $m_z=\av{\sigma_z}$
for the spin in the center of the block, and consider the effect of
different block sizes compared to Onsager's exact
solution\cc{Onsager1944,Yang1952} (continuous line). We see that
with block size $5\times 5$ we already obtain very accurate results
away from the critical point (with $\beta_c\approx 0.44$), and
increasing the block size, we obtain more accurate results near
$\beta_c$. Here we note that our results close to $\beta_c$ are not
as accurate as those obtained in Refs.\cc{LevinNave2007,
XieXiang2009, ZhaoXiang2010, XieXiang2012, EvenblyVidal2015},
however, they are more accurate than the results obtained using
other variants of the BP algorithms such as in
Refs.\cc{ZhouWang2012, WangZhou2013, ZhouZheng2015}. 

In \Fig{fig:figS4}(b), we used blockBP + ITE to compute the local
magnetization $m_z=\av{\sigma_z}$ at the ground state of the
infinite transverse Ising model for different values of the external
field $B$, and compared our results to those obtained by the corner
transfer matrix (CTM) from \cRef{PhienOrus2015}. We set the center
size to be $2\times 2$, and use a block size $4\times 4$. To compute
$m_z$ for the final converged infinite system, we have copied the
block into a $52\times 52$ finite tensor network and computed the
average $m_z$ at the central $2\times 2$ cell using bMPS. We can see
that away from criticality ($B_c\approx 3.044$\cc{BloteDeng2002}),
our results agree well with the corner transfer matrix results,
especially for $D=3$. For $D=2$, our results approach better the
results at $D=3$ for $B < 3.1$.  We note that with $4\times 4$ block
size our simulation is extremely efficient ($\approx 0.4$s and
$\approx 1.1$s per iteration for $D=2$ and $D=3$ respectively using
a single core of $2.3$ GHz frequency). As in the classical Ising
model, with a small block size we are not able to accurately
reproduce $m_z$ near $B_c$. Nevertheless, the results close to the
phase transition can be systematically improved by increasing the
block size.

%
%

\section{Summary and outlook}

In summary, we have proposed a belief propagation algorithm for
approximately contracting 2D tensor networks. Our approach is
straightforwardly applicable for both finite and infinite systems,
different unit cell sizes, and can be readily generalized to
different geometries. Furthermore, our method allows straightforward
and efficient parallelization. The accuracy and efficiency of the
method are demonstrated with applications to prototypical quantum 2D
systems, and benchmarked with state-of-the-art results based on
imaginary time evolution. 
Future directions of investigation include testing the
algorithm on different lattices, including higher dimensions and/or
different geometries, as well as using gradient-based methods to
optimize the PEPS instead of ITE\cc{SchebNoack2023}, as done for
CTM\cc{Corboz2016,LiaoXiang2019}.

%
%

\acknowledgements 
  We are grateful to J. Hasik for his comments on a previous version
  of paper, and for providing us data of iPEPS ground state (with
  $D=3$) for critical transverse Ising model obtained from
  gradient-based optimization, based on which we verified that with
  blockBP we need a large block size ($100\times 100$) as well as a
  larger $\chi_m$ ($\chi_m=25$) to reach similar precision. 
  C. G. acknowledges support from the Open Research Fund from State Key Laboratory of High Performance Computing of China (Grant No. 202201-00).
  I.A.\ acknowledges the support of
  the Israel Science Foundation (ISF) under the Individual Research
  Grant No.~1778/17 and joint Israel-Singapore NRF-ISF Research
  Grant No.~3528/20. D.P. acknowledges support from joint
  Israel-Singapore NRF-ISF Research grant NRF2020-NRF-ISF004-3528.

\bibliographystyle{apsrev4-2}
\bibliography{refs}

\appendix

\section{Details of the blockBP implementation}\label{sec:zipup}

The two building blocks of our blockBP algorithm are 1) computing
the fixed points of the MPS messages and 2) performing computations
(either computing local observables or updating the site tensors)
inside the center of each block. We use the boundary MPS (bMPS)
\cRef{LubaschBanuls2014b} algorithm for both tasks, however, it is
important to stress that in our case the bMPS algorithm is only
applied to a small portion of the system, our blocks, instead of the
entire system, thus making our computations much more manageable and
also parallelizable.

Let us consider, for example, the case shown in Fig.1(d) of the main
text. Here one wants to evaluate the message going from right to
left from a certain block. We then need to contract the local tensor
with the three incoming messages from the top, the right and the
bottom. Each of these messages is represented by an MPS. The
contraction of the messages with the tensors in the block is done
from right to left, column by column, each time keeping a bond
dimension of (at most) $\chi_m=D^2$, where $D$ is the virtual bond
dimension of the tensors in the block. If the block we are
considering is at the boundary of a finite system, then the MPS
messages going into and out of it consist of only trivial tensors.

Once the messages have converged, we can compute an observable
within a block, or update the local site tensors inside a block
during the imaginary time evolution. Following Fig.1(f) of the main
text, we need to evaluate the contraction between the four converged
incoming MPS messages, from top, bottom, left and right, with the
tensors of the block (note that this can be straightforwardly
generalized to other 2D lattices). Also in this case one uses the
bMPS algorithm on a single block plus messages, from any of the
sides of the block. Since we aim to be more accurate when evaluating
local observable at the center of the block, we use a larger bond
dimension $\chi=2D^2+10$ when doing this contractions.

\begin{figure}
  \includegraphics[width=\columnwidth]{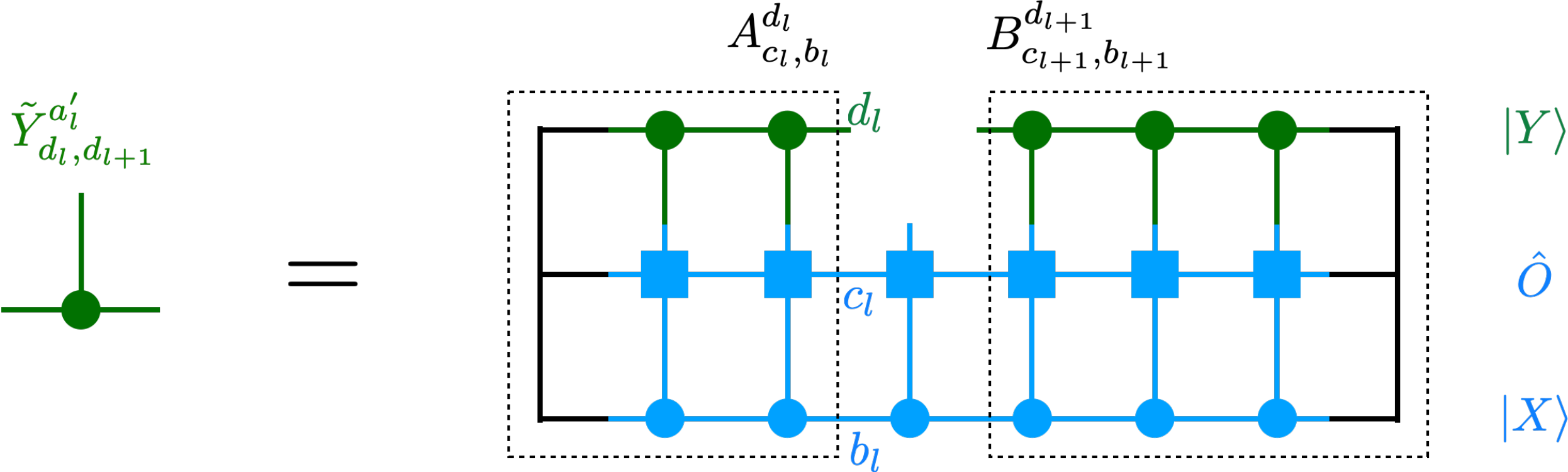}
    \caption{Updating a local site tensor in the zip-up algorithm to
    multiply an MPS by an MPO, which corresponds to
    Eq.(\ref{eq:update}). The left block is represented by the
    tensor $A^{d_l}_{c_l,b_l}$, while the right block by
    $B^{d_{l_+1}}_{c_{l+1},b_{l+1}} $. The tensor
    $\tilde{Y}^{a_l'}_{d_l,d_{l+1}}$ is obtained contracting these
    two tensors with $O^{a_l', a_l}_{c_l, c_{l+1}}$ and $X_{b_l,
    b_{l+1}}^{a_l}$. } \label{fig:figS1}
\end{figure}

The most computationally intensive subroutine involved in both tasks
in the case of a square lattice is to multiply one line of
$3-$legged tensors which represent an (updated) environment, with a
line of $4-$legged tensors, e.g. the tensor of the block next to the
environment. While this can be done by using the standard MPO time
MPS arithmetic~\cite{Schollwock2011}, and then compressing the
resulting MPS, we use the following zip-up algorithm that fuses
these two operations into one to reduce both the memory and
computational cost, as shown in \cite{VerstraeteMurg2008}.

We can write the input MPS $\ket{X}$ as
\begin{align}
  \ket{X} = \sum_{j=1}^L X^{a_1}_{b_1, b_2} 
    X^{a_2}_{b_2, b_3} \cdots X^{a_L}_{b_L, b_{L+1}},
\end{align}
and the input MPO $\hat{O}$ as
\begin{align}
  \hat{O} = \sum_{j=1}^L O^{a_1', a_1}_{c_1, c_2} 
    O^{a_2', a_2}_{c_2, c_3} \cdots O^{a_L', a_L}_{c_L, c_{L+1}}.
\end{align}
Our task is to calculate a fixed bond dimension MPS $\ket{Y}$, which
approximates $\hat{O}\ket{X}$. We begin by first initializing
$\ket{Y}$ randomly:
\begin{align} 
  \ket{Y} = \sum_{j=1}^L Y^{a_1'}_{d_1, d_2} Y^{a_2'}_{d_2, d_3} 
    \cdots Y^{a_L'}_{d_L, d_{L+1}}. 
\end{align}
Then we use an iterative algorithm to minimize the distance
\begin{align}\label{eq:distance}
  &\norm{\ket{Y} - \hat{O}\ket{X}}^2 \nonumber \\ 
  =& \braket{Y}{Y}
    - \bra{Y}\hat{O}\ket{X} - \bra{X}\hat{O}^\dagger\ket{Y}
    + \bra{X}\hat{O}^{\dagger}\hat{O}\ket{X}.
\end{align}
Note that the last term on the right hand side of \Eq{eq:distance}
is a constant and can be neglected during the optimization. To
optimize $\ket{Y}$, we iteratively update each site tensor of it
using DMRG-like sweeps. For a specific site $l$, we first compute
\begin{align}\label{eq:update}
  \tilde{Y}^{a_l'}_{d_l, d_{l+1}} 
    = \sum_{b_l, c_l, b_{l+1}, c_{l+1}, a_l} 
    A^{d_l}_{c_l, b_l} O^{a_l', a_l}_{c_l, c_{l+1}} 
    B^{d_{l+1}}_{c_{l+1}, b_{l+1}} X^{a_l}_{b_l, b_{l+1}},
\end{align}
where the rank-$3$ tensors $A$ and $B$ can be computed iteratively
using
\begin{align}
  A^{d_l}_{c_l, b_l} = \sum_{b_{l-1}, c_{l-1}, d_{l-1}, 
    a_{l-1}, a_{l-1}'} &A^{d_{l-1}}_{c_{l-1}, b_{l-1}} 
    Y^{a_{l-1}'}_{d_{l-1},d_{l}} \times \nonumber \\ 
    &O^{a_{l-1}',a_{l-1}}_{c_{l-1},c_l} 
    X^{a_{l-1}}_{b_{l-1}, b_l}
\end{align}
and 
\begin{align}
  B^{d_{l+1}}_{c_{l+1}, b_{l+1}} 
    = \sum_{b_{l+2}, c_{l+2}, d_{l+2}, a_{l+1}, a_{l+1}'} 
      &B^{d_{l+2}}_{c_{l+2}, b_{l+2}}Y^{a_{l+1}'}_{d_{l+1},d_{l+2}} \times
      \nonumber \\ 
      &O^{a_{l+1}', a_{l+1}}_{c_{l+1}, c_{l+2}} 
      X^{a_{l+1}}_{b_{l+1}, b_{l+2}}
\end{align}
with $A^{d_1}_{c_1, b_1} = B^{d_{L+1}}_{c_{L+1}, b_{L+1}} = 1$.
During the left to right sweep, we perform a QR decomposition of
$\tilde{Y}^{a_l'}_{d_l, d_{l+1}}$ and get
\begin{align}
  \mathrm{QR}(\tilde{Y}^{a_l'}_{d_l, d_{l+1}}) 
    = \sum_s Q^{a_l'}_{d_l, s} R_{s, d_{l+1}},
\end{align}
and then we take $Q^{a_l'}_{d_l, s}$ as the new site tensor. During
the right to left sweep, we perform an LQ decomposition of
$\tilde{Y}^{a_l'}_{d_l, d_{l+1}}$ and get
\begin{align}
  \mathrm{LQ}(\tilde{Y}^{a_l'}_{d_l, d_{l+1}}) 
    = \sum_s L_{d_l, s} Q^{a_l'}_{s, d_{l+1} },
\end{align}
and then we take $Q^{a_l'}_{s, d_{l+1} }$ as the new site tensor.
\Eq{eq:update} is also demonstrated in \Fig{fig:figS1}. In addition,
from \Eq{eq:update} we have $-\|\tilde{Y}^{a_l'}_{d_l, d_{l+1}}\|^2
= \|\;\ket{Y} - \hat{O} \ket{X} \|^2 - \bra{X}
\hat{O}^\dagger\hat{O}\ket{X}$, which is exactly the loss function
subtracted by a constant term, therefore it can be used to monitor
the convergence, similar to the ground state energy in DMRG
\cite{Schollwock2011}.  In our simulations we have set the
convergence criterion to be the standard deviation of
$-\|\tilde{Y}^{a_l'}_{d_l, d_{l+1}}\|^2$ for a full sweep (left to
right and then right to left) divided by the mean value, with a
tolerance of $10^{-6}$ and the maximum number of sweeps to be $10$. 

For computing the MPS messages of a double layer tensor network, the
MPS messages are randomly initialized as a matrix product density
operator\cc{VerstraeteCirac2004} using normal distribution, while
for single layer tensor network which is used for classical models
(see the simulation of the classical Ising model below), the MPS
messages are simply initialized as a random MPS using uniform
distribution. Assuming that the mean square error between the MPS
messages in $l$-th step and those in the $l-1$-step is $\epsilon_l$,
the convergence criterion in this case is chosen to be $\epsilon_l /
\epsilon_1$, with a tolerance of $10^{-5}$ and the maximum number of
iterations to be $10$.

\section{The imaginary time evolution algorithm for PEPS}\label{sec:ite}

\begin{figure}
  \includegraphics[width=\columnwidth]{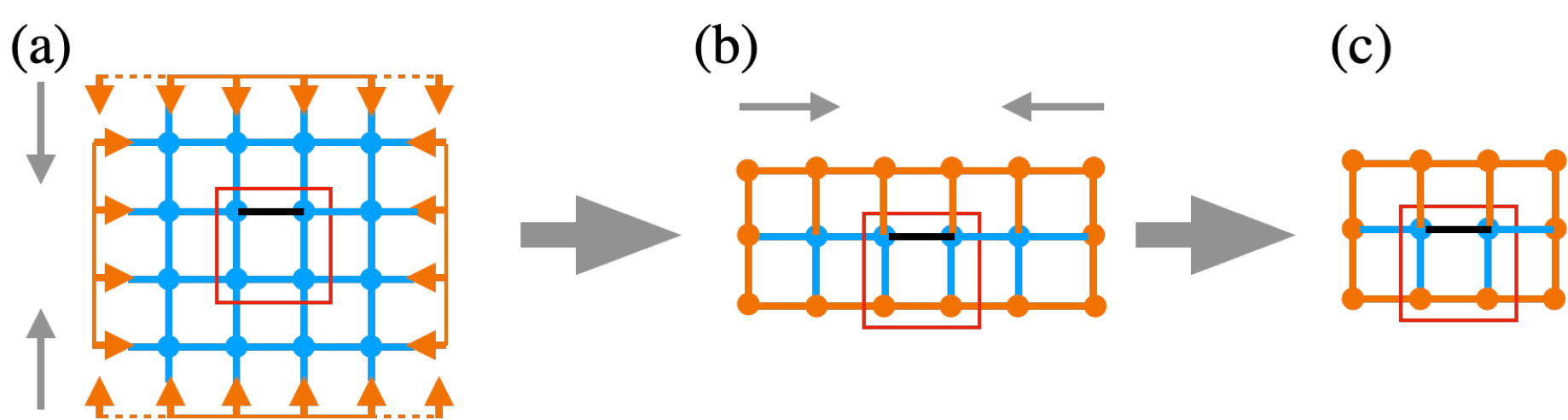}
  \caption{Procedures to compute the local environment for the black
  bond inside a specific block taken from Fig.1(e) of the main text.
  The standard bMPS algorithm for finite PEPS is used for this
  calculation.} \label{fig:bmpsupdate}
\end{figure}

Generally, two different approaches are used to optimize the PEPS
tensors for computing the ground state, namely variational
minimization and imaginary time evolution. In this work we focus on
the second approach. For a 2D Hamiltonian on a square lattice,
written as $\Hop = \sum_i h_i$, where $i$ stands for all the nearest
neighbor pairs of sites (bonds) and $h_i$ denotes the local
Hamiltonian term on the $i$-th bond, we first perform a first-order
expansion of the imaginary time evolutionary operator $U(d\tau) =
e^{-\Hop d\tau}$ for a small time interval $d\tau$ as
\begin{align}
\label{eq:Ut}
  U(d\tau) \approx e^{-h_1 d\tau} e^{-h_2 d\tau} 
    e^{-h_3 d\tau}  \cdots .
\end{align}
Each term on the right hand of \Eq{eq:Ut} acts on a specific bond,
referred to as a gate. Changing the order of these terms on the
right hand of \Eq{eq:Ut} still results in a first order
approximation, therefore one could freely adjust the order to
maximally reuse the intermediate computations during the time
evolution. In our implementation we will first apply the gates on
the horizontal bonds row by row, and then apply the gates on the
vertical bonds column by column. 
 
The application of a specific gate $G_i = e^{-h_i d\tau}$ is
formulated as an optimization problem to minimize the loss function
$\norm{\ket{\psi'} - G_i \ket{\psi}}^2$, where $\ket{\psi'}$ is the
new PEPS with all the tensors the same as $\ket{\psi}$ except the
two tensors at bond $i$ and $\norm{\cdot}$ means the $2$-norm.
Performing the local optimization requires to calculate the
environment around the tensors being updated, which is the central
difference among various PEPS updating algorithms. In our approach,
we first use blockBP to compute the boundary messages as matrix
product states (MPSs) for each block, which plays the role of the
environment for each block.  Then performing imaginary time
evolution on the tensors inside each block becomes almost the same
as the original PEPS updating problem, except that we are dealing
with a much smaller finite PEPS, and that there are four nontrivial
MPS on the boundaries (the standard finite PEPS update can also be
taken as a special case where the boundaries are trivial MPSs for
which the site tensors only contain a single element $1$). As such
we can use existing finite PEPS update algorithms to update the
tensors inside each block. In our implementation we use the boundary
MPS (bMPS) method algorithm inside each block, as demonstrated in
Fig.~\ref{fig:bmpsupdate} for applying a single gate on the black
bond in a block taken from Fig.1(e) of the main text. In
Fig.~\ref{fig:bmpsupdate}(a), we map our problem to be exactly the
same as the standard finite PEPS problem by padding four trivial
tensors with single element $1$ at the four corners. Then we treat
the top and bottom rows of tensors as two MPSs, the middle rows of
tensors as matrix product operators (MPOs), and apply the MPOs onto
the boundary MPSs (see Appendix.~\ref{sec:zipup} for the MPO-MPS
multiplication) until there are only three rows left, as shown in
Fig.~\ref{fig:bmpsupdate}(b). After that, we perform tensor
contractions from left and right until only the minimal environment
for the two tensors is left, as shown in
Fig.~\ref{fig:bmpsupdate}(c). Finally, we perform the alternating
least squares scheme, in which one of the two tensors will be
updated at a time while keep the other as constant, until the final
loss converges. Overall, our implementation of bMPS update for each
block closely follows the algorithm in
Ref.~\cite{LubaschBanuls2014}.

\section{Evaluation of reduced density matrix for finite systems}\label{sec:rdm}

\begin{figure}
  \includegraphics[width=0.9\columnwidth]{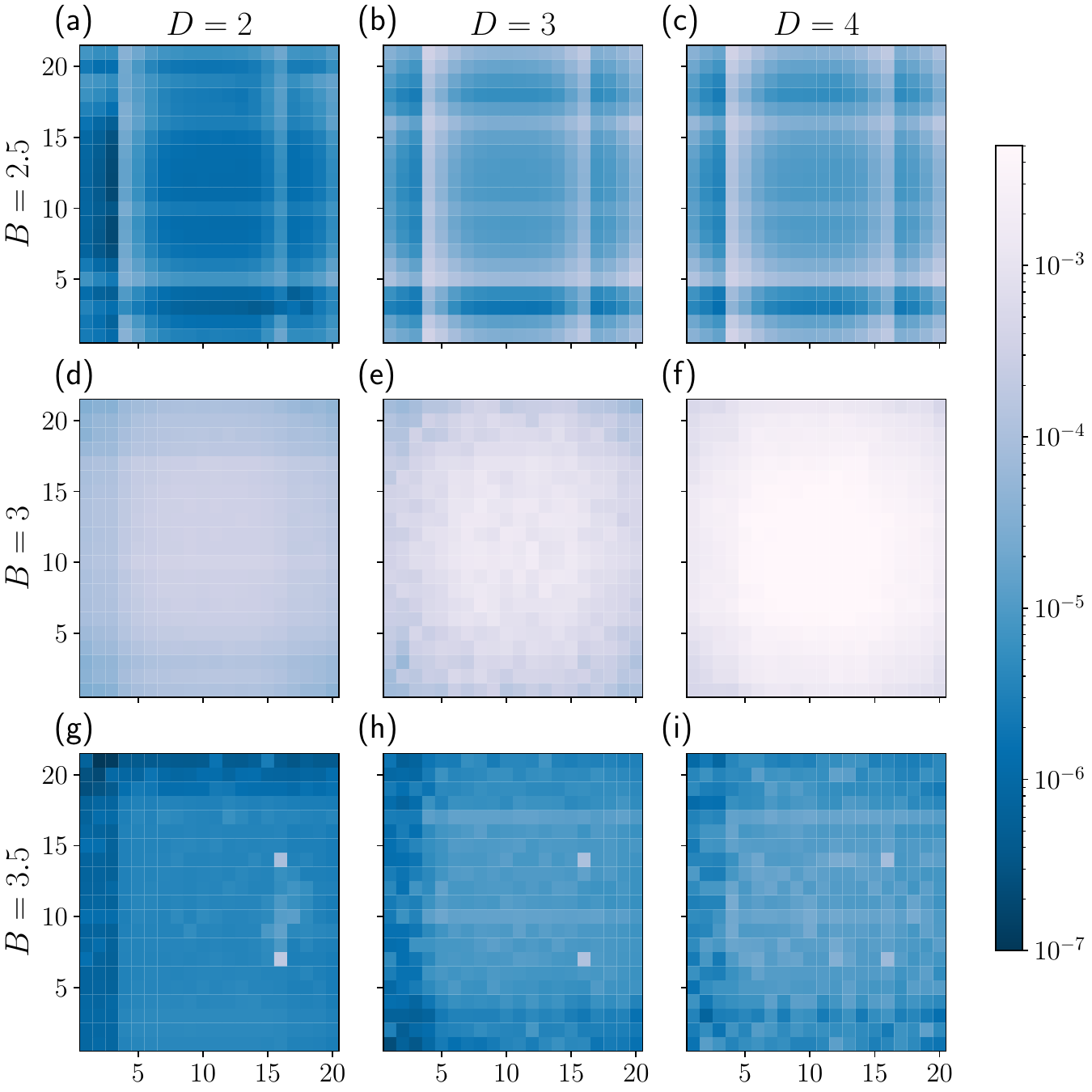}
  \caption{Trace distances between the single site reduced density
  matrices of the horizontal bonds computed using bMPS and blockBP
  with block size $7\times 7$. The ground states of the $21\times
  21$ transverse Ising model for different local fields $B$ are used
  to compute the reduced density matrices.  The columns from left to
  right are results for $D=2,3,4$ respectively (i.e. panels
  (a,d,g) correspond to $D=2$ and so on), while the rows from top
  to bottom are results for $B=2.5,3.0,3.5$ respectively (i.e.
  panels (a,b,c) are for $B=2.5$). }
  \label{fig:figS2}
\end{figure}

\begin{figure}
  \includegraphics[width=0.9\columnwidth]{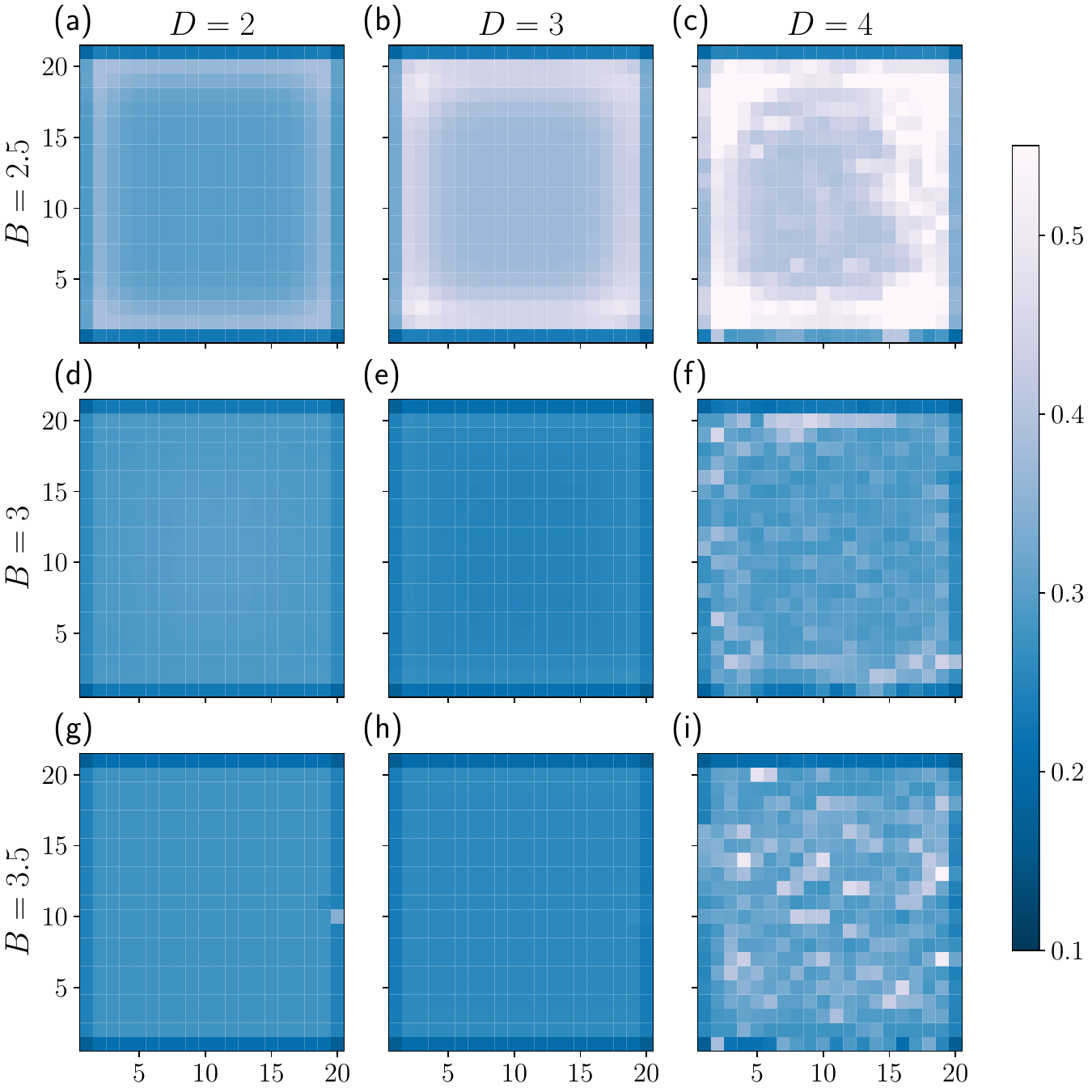}
  \caption{Trace distances between the single site reduced density
  matrices of the horizontal bonds computed using bMPS and simple
  update. The other settings are the same as Fig.~\ref{fig:figS2}. }
  \label{fig:figS2u}
\end{figure}

Here we study the quality of the local environments in our blockBP
by computing the two-site reduced density matrices on each
horizontal bond using blockBP and compare them to the reduced
density matrices computed using bMPS.  The comparison is shown in
the form of a heat map in \Fig{fig:figS2}, where we consider the
ground state of a $21\times 21$ transverse Ising model with OBC. 
The blockBP reduced density matrices were computed using $7\times 7$
blocks.  Quantitatively, what we do is to compute the reduced
density matrix for the site in row $i$ and columns $j$ and $j+1$
with both methods, i.e. $\rho_{i,j,j+1}^\text{bMPS}$ and
$\rho_{i,j,j+1}^\text{blockBP}$, and compute the trace distance
between them $d =
\Tr(|\rho_{i,j,j+1}^\text{blockBP}-\rho_{i,j,j+1}^\text{bMPS}|)/2$
with $|Z| = \sqrt{Z^{\dagger}Z}$.  We can see that for $B=2.5,3.5$,
which are away from the critical value ($B_c\approx
3.044$~\cite{BloteDeng2002}), the reduced density matrices computed
by these two methods are very close to each other with average
distance lower than $10^{-5}$ for all the bond dimensions $D$
considered, while for $B=3.0$ we get $\distance \approx 10^{-3}$ on
average.  This indicates that bMPS and blockBP lead to quantitative
similar results, especially away from the transition point.

For comparison, in \Fig{fig:figS2u} we used the same setting as in
\Fig{fig:figS2}, and calculated the trace distance between the
reduced density matrices computed using simple-update (or, in other
words, the plain BP algorithm where the message MPS has only one
leg) and the reduced density matrices computed using bMPS. As can be
seen from these heat maps, in these cases, the trace distance is of
the order $10^{-1}$ showing that blockBP environments are far
superior to those of the plain BP, or equivalently, the
simple-update algorithm.

\section{Convergence of blockBP for finite systems}\label{sec:convergence}

\begin{figure}
  \includegraphics[width=\columnwidth]{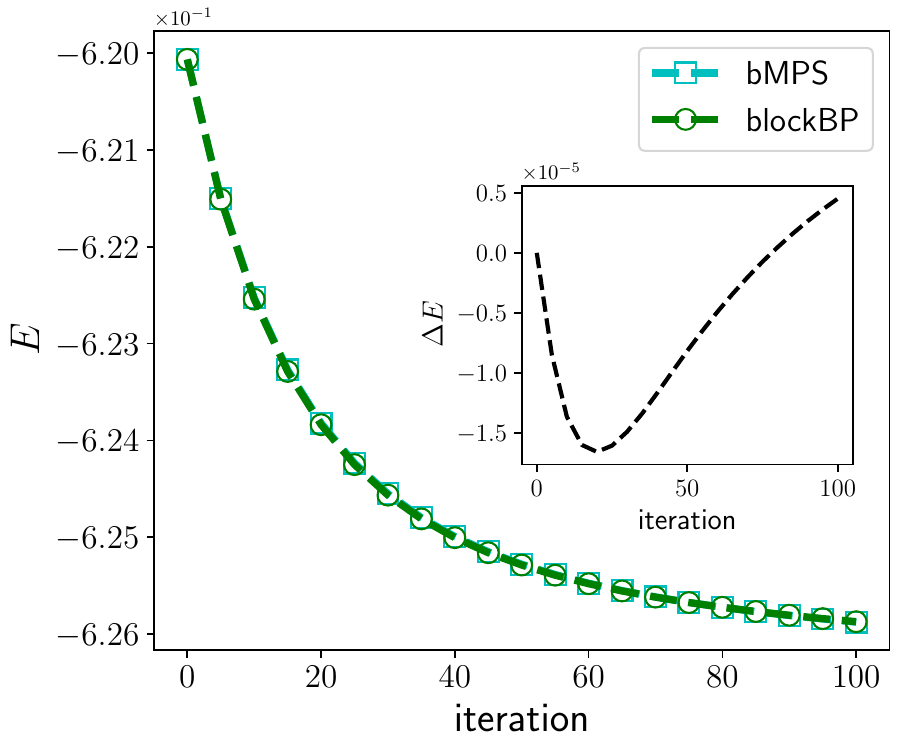}
  \caption{The ground state energy $E$ obtained using bMPS (light
  blue squares) and blockBP (green circles) as a function of the
  imaginary time evolution steps. The inset shows the difference
  between the energies computed for the two algorithms, namely
  $\Delta E = E_{{\rm blockBP}} - E_{{\rm bMPS}}$, the initial state
  for both cases is chosen as the ground state of the $10\times 10$
  AFH model with $D=3$ and we have used $d\tau=0.01$ in both cases.
  } \label{fig:figS3}
\end{figure}

In the main text we have benchmarked the imaginary time evolution algorithm with blockBP
update against the same algorithm with bMPS full update. We
considered the anti-ferromagnetic Heisenberg (AFH) model and looked
at the final result of the energy.  Here we further demonstrate that
even the evolution of our blockBP update towards the ground state
has a very similar dynamics as bMPS, given the same parameters and
the same initial state. This is due to the fact that both algorithm
use the same Trotterized imaginary time evolution
\cite{HatanoSuzuki2005} and they estimate in a very similar manner
the Trotter steps.  Specifically, in \Fig{fig:figS3} we consider the
imaginary time evolution of a $10\times 10$ AFH model using both
bMPS and blockBP update (with block size $5\times 5$) with
$d\tau=0.01$ and $D=4$. For both algorithms we took as initial
condition the ground state obtained using bMPS for $D=3$.  We can
see from the inset that the difference between bMPS and blockBP is
around $10^{-5}$ throughout the imaginary time evolution.

\section{Computing local observables of infinite system using blockBP}\label{sec:expec}

\begin{figure*}
\includegraphics[width=2\columnwidth]{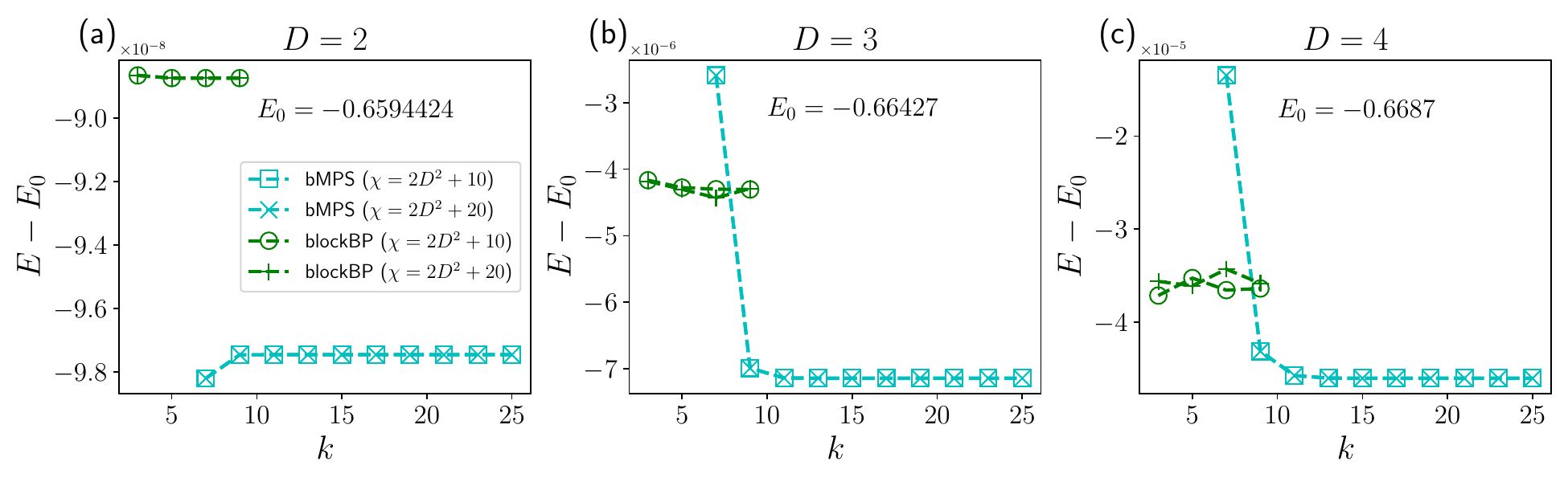}
  \caption{Ground state energies of the infinite AFH model as a
  function of $k$ computed using bMPS and blockBP for (a) $D=2$, (b)
  $D=3$ and (c) $D=4$. For all those panels the cyan dashed lines
  with squares and x are energies computed using bMPS with
  $\chi=2D^2+10$ and $\chi=2D^2+20$ respectively, while the green
  dashed lines with circle and + are energies computed using blockBP
  with $\chi=2D^2+10$ and $\chi=2D^2+20$ respectively.  }
  \label{fig:figS5}
\end{figure*}

In the following, we show the precision of our blockBP when used for
computing local observables of an infinite system. Concretely,
taking the infinite tensor network ground state computed in Fig.~4
in the main text using blockBP for different bond dimensions $D$, we
compared two different ways of computing its energy. In the first
approach, we embedded the tensor network with central block of size
$2\times 2$ into a finite system of size $(2k+2)\times (2k+2)$ where
the boundaries are randomly initialized, and then we compute the
energy in the $2\times 2$ center using bMPS. Since the the system is
infinite, by embedded we mean that the blocks are repeated to form
the system. In the second approach, we calculated the ground energy
using blockBP with a $2k\times 2k$ block size. In this case, the
boundary of the system are given by the MPS messages, while for bMPS
we have used a random MPS. The results are shown in \Fig{fig:figS5}.
Given the high accuracy of the results, we show both for blockBP and
bMPS the energy minus an offset value of $E_0$. We see that both
approaches converge to the $5$-th digit for large enough $k$, but
that the blockBP results converge much faster --- already with a
small $k=3$. Therefore blockBP can also be used as an efficient
method to accurately compute local observables in infinite systems.
Fig.~\ref{fig:figS5} also shows the ground state energies computed
using a larger $\chi=2D^2+20$, and as we can see, they are very
close to the energies computed using $\chi=2D^2+10$ (the latter is
chosen as the default throughout this work).


\begin{figure*}
\includegraphics[width=2\columnwidth]{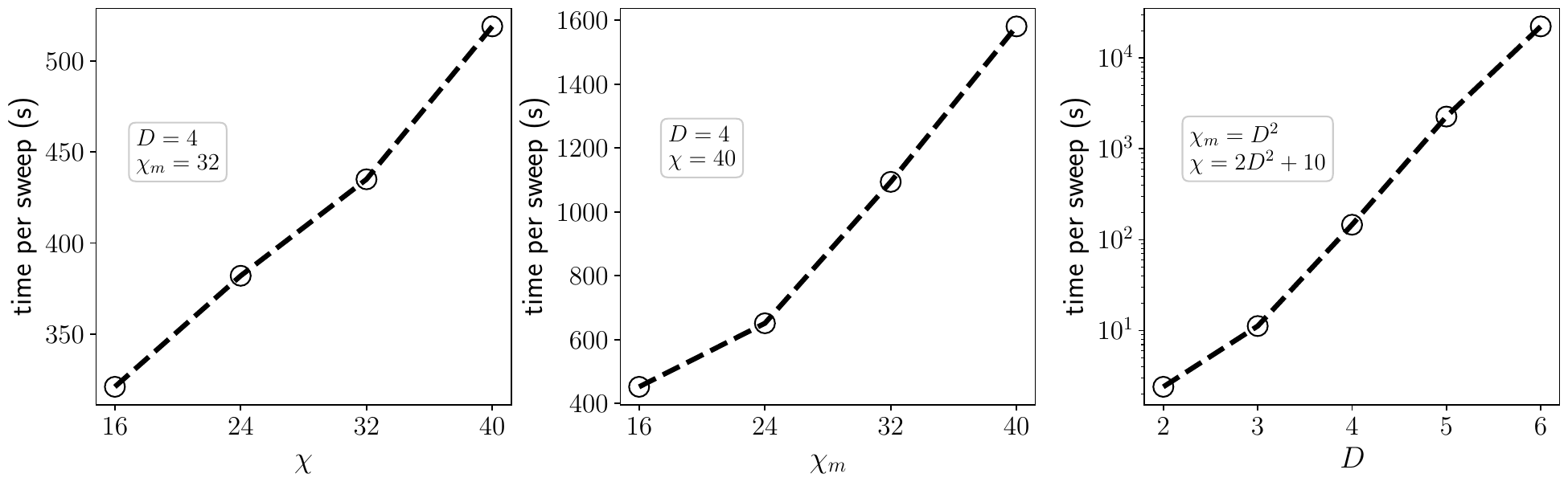}
  \caption{Runtime per ITE sweep of blockBP for finite AFH model
  with size $14\times 14$ against the hyperparameters $\chi$ (a),
  $\chi_m$ (b) and $D$ (c). We have used a block size of $7\times
  7$. A single thread of the Intel(R) Core(TM) i7-11850H CPU (2.5
  GHz) is used. } \label{fig:figS7}
\end{figure*}

\section{Runtime scaling of blockBP against
hyperparameters}\label{sec:scaling}

\begin{figure*}
\includegraphics[width=2\columnwidth]{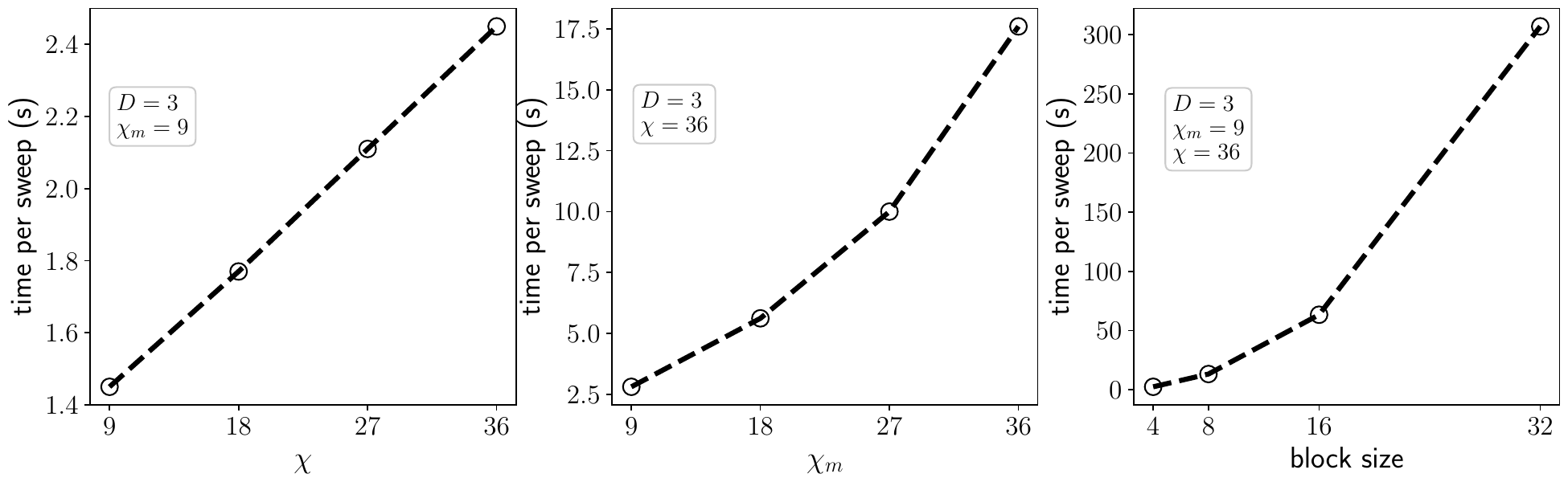}
  \caption{Runtime per ITE sweep of blockBP for infinite transverse
  Ising model against the hyperparameters $\chi$ (a), $\chi_m$ (b)
  and block size (c). In (a,b) we have used a block size of $4\times
  4$, we have also used $D=3$ for all the panels. A single thread of
  the Intel(R) Core(TM) i7-11850H CPU (2.5 GHz) is used. }
  \label{fig:figS8}
\end{figure*}

\begin{figure}
\includegraphics[width=\columnwidth]{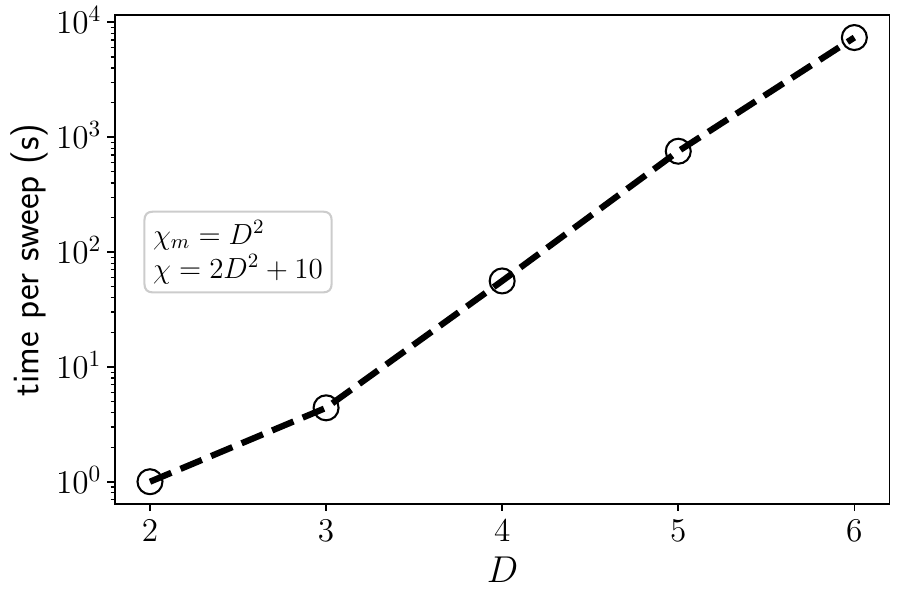}
  \caption{Runtime for $5$ ITE sweeps of blockBP for infinite AFH
  model against $D$. We have used a block size of $4\times 4$. $8$
  threads of the Intel(R) Core(TM) i7-11850H CPU (2.5 GHz) are
  used.} \label{fig:figS9}
\end{figure}

Here we show more details of the runtime scaling of our blockBP
versus the $4$ hyperparameters, namely the bond dimension $D$ of the
PEPS, the bond dimension $\chi$ used during bMPS update, the bond
dimension $\chi_m$ used for computing the MPS messages, and the
block size. Specifically, in Fig.~\ref{fig:figS7}, we show the
runtime scaling of blockBP for finite AFH model against $\chi$,
$\chi_m$ and $D$ with block size fixed to be $7\times 7$. In
Fig.~\ref{fig:figS8}, we show the runtime scaling of blockBP for
infinite transverse Ising model against $\chi$, $\chi_m$ and block
size with $D$ fixed to be $3$. In Fig.~\ref{fig:figS9} we show the
runtime scaling of blockBP for infinite AFH model against $D$, with
$\chi$ and $\chi_m$ using their default values as in the main text.
From Figs.~(\ref{fig:figS7}, \ref{fig:figS8}) we can see that the
runtime scales extremely rapidly with $D$, and scales slightly
faster than linear against $\chi_m$ and the block size, while in
comparison the scaling against $\chi$ is a lot slower than $\chi_m$.
The latter is due to the fact that when computing messages we have
used several iterations, while during the bMPS update performed in
each block, only two iterations are required (one horizontal and one
vertical). Nevertheless, as shown in the main text, the default
value of $\chi_m = D^2$ could already result in accurate results in
all the cases considered in this work. In Fig.~\ref{fig:figS9} we
show the runtime scaling of the infinite AFH model against $D$, with
$\chi$ and $\chi_m$ set to their default values. We note that the
runtime scaling of a fast full update algorithm has been studied in
Fig.13 of Ref.~\cite{PhienOrus2015}, where we can see that for
$D\leq 5$ we are faster while for $D=6$ we are slower. However we
stress that this comparison is not exact since the computational resources used for Fig.13 
of Ref.~\cite{PhienOrus2015} are not shown in that paper.

\end{document}